# Manager Characteristics and SMEs Restructuring Decisions: In-Court vs. Out-of-Court Restructuring

**Rachid ACHBAH**

University Lyon 2

Coactis-EA 4161

14 avenue Berthelot 69363 Lyon, France

rachid.achbah@univ-lyon2.fr

**Abstract**

This study aims to empirically investigate the impact of managers' characteristics on their choice between in-court and out-of-court restructuring. Based on the theory of upper echelons, we tested the preferences of 342 managers of financially distressed French firms regarding restructuring decisions. The overall findings of this study provide empirical support for the upper echelons theory. Specifically, managers with a long tenure and those with a high level of education are less likely to restructure before the court and are more likely to restructure privately. The findings also indicate that managers' age and gender do not significantly affect their choice between in-court and out-of-court restructuring. This study contributes to the literature on bankruptcy and corporate restructuring by turning the focus from firm characteristics to manager characteristics to explain restructuring decisions.

# INTRODUCTION

Managers' characteristics matter at every stage of the organizational life cycle. Building on upper echelons theory (Hambrick and Mason, 1984), an extensive body of literature has studied the effects of top managers' characteristics on firm leadership, strategic orientation, and outcomes, as they play an essential role in the firm's decision process (e.g., Elbanna, Thanos, and Jansen, 2020; Fuming, Subramaniam, and Jizhong, 2018; Liu, Fisher, and Chen, 2018; Khelil, 2016; Walayet and Vieito, 2013). This literature focuses on managers' characteristics with two underlying assumptions. First, critical decision-making power is centered on executives' hands, especially in the context of small and medium enterprises (Finkelstein and Hambrick, 1996; Khelil, 2016; Zemis and Demil, 2020). Second, managers have different personal perspectives that are influenced by their personalities, values, and experiences (Finkelstein, Hambrick, and Cannella, 2009). The main concern in this line of research is to explain the strategic choices made by managers based on their characteristics, which have been studied in various streams of the literature, ranging from corporate finance to strategic management. See Busenbark, Krause, Boivie, and Graffin (2016) for more detailed literature reviews.

A large majority of this literature focuses on the effect of individual characteristics on functional or growth-oriented decisions such as investments, performance, and innovation[1]. By contrast, scholars have paid less attention to how managers' characteristics influence firm-level decisions in times of hardship. For example, the literature has overlooked the important question of how managers with different characteristics react and reorganize in financial distress.

In this paper, we examine the effects of four top manager characteristics in times of financial distress: age, gender, educational level, and tenure. We define financial distress in its most general specification, i.e., a situation where the firm cannot meet its current debt obligations and has a significantly decreased distance to default (Blazy, Martel, and Nirjhar, 2014). In this specific context, the manager plays a crucial role in designing and implementing a strategy called a "restructuring process," which will hopefully allow the firm to transition toward recovery (Koh, Durand, Dai, and Chang, 2015). Hence, restructuring is a form of corpo-rate reorganization encompassing strategic, financial, and organizational practices aimed at restoring a financially distressed firm's health (Bowman and Singh, 1993; Girod and Whittington, 2017). One of the most critical decisions in the restructuring process is the choice of the restructuring method. Financially distressed firms may choose to restructure the firm privately - by initiating procedures such as negotiations with creditors, cost-cutting strategies, and asset liquidations, or they may choose an in-court restructuring by initiating insolvency proceedings (Jostarndt and Sautner, 2010; Blazy et al., 2014; Fisher, Martel, and Naranjo, 2022). The costs and benefits of the in-court and out-of-court dichotomy have been explored by existing literature (Gilson, John, and Lang, 1990; Blazy et al., 2014; Jostarndt and Sautner, 2010), and several firm-level predictor variables have been identified (Blazy et al., 2014; Jostarndt and Sautner, 2010). However, previous literature has not yet investigated how the characteristics of the managers of distressed firms influence their choices regarding restructuring decisions, especially the choice between in-court and out-of-court restructuring.

---

[1]Specifically top manager's characteristics have been shown to influence: Decision making (Fuming Jiang et al., 2018), firm performance (Kaur and Singh, 2019; Liu, Fisher, and Chen, 2018; Walayet and Vieito, 2013), growth decisions and mergers and acquisition (Malmendier and Tate, 2008; Aktas et al., 2016), corporate investment (Malmendier and Tate, 2005) and many other key decisions in the corporate lifecycle.

This study directly addresses how managerial characteristics affect decision-making in times of adversity by exploring one of the most crucial decisions that top managers have to make within the specific context of financial distress. The theoretical framework integrates two strands of existing research. On the one hand, we refer to the upper echelons theory (Hambrick and Mason, 1984), which claims that the managers' characteristics determine how they perceive, interpret, and react to environmental stimuli and, therefore, influence their strategic choices (Chatterjee and Hambrick, 2007; Gordon, Karel, Johnny, and Xin, 2021; Campbell, Jeong, and Graffin, 2019; Liu et al., 2018; Schumacher, Keck, and Tang, 2020; Marquez-Illescas, Zebedee, and Zhou, 2019). On the other hand, we build upon the literature on corporate restructuring and reorganization (Blazy et al., 2014; Adriaanse and Rest, 2017; Jostarndt and Sautner, 2010; Hotchkiss, Mooradian and Thornburn, 2008) to identify the contextual constraints under which managers choose their restructuring strategy and the way each attribute is likely to influence their choice between in-court and out-of-court restructuring. By integrating these two strands, we develop a theoretical model to understand how managers' characteristics shape their choices between in-court and out-of-court restructuring.

To investigate this question empirically, we conducted tests on an original sample collected via a survey. The survey was administered directly to managers of French small and medium enterprises that qualified as financially distressed. Our findings revealed some interesting results. We provide evidence that managers' sociodemographic characteristics, including age and gender, do not significantly affect their choice between in-court and out-of-court restructuring. We found that manager with a higher tenure is less likely to prefer in-court re-structuring and more likely to choose private restructuring. Additionally, we show that a manager's educational level decreases the likelihood of in-court restructuring. More precisely, the higher the educational level, the lower the probability of in-court restructuring, and the higher the likelihood of private restructuring.

To answer this question, we begin by reviewing the literature on financial distress resolution. We continue by contrasting the two main options available to managers based on their perceptions and orientations. This analysis and contrast between the two main restructuring approaches allow us to see the extent to which demographic variables, tenure, and education level more or less strongly orient the executive toward a choice. We then propose hypotheses about the possible effects of the main variables on the choice of restructuring approach.

Our study contributes to at least three strands of the literature. First, it contributes to the literature on corporate restructuring. Previous studies have examined the choice between in-court and out-of-court restructuring exclusively based on firm characteristics (Gilson et al., 1990; Peek and Rosengren, 2005; James, 1995; Blazy et al., 2014). However, to the best of our knowledge, our study is one of the few to provide evidence that managerial characteristics affect the choice between in-court and out-of-court restructuring. Specifically, this study focuses on managers rather than firm characteristics in terms of corporate restructuring. Second, regarding upper echelons theory, previous research has examined the critical role of the executive's attributes and characteristics in shaping organizational outcomes and strategic decisions (Gordon et al., 2021; Campbell et al., 2019; Schumacher et al., 2020; Benischke, Martin, and Glaser, 2019). We contribute to this literature by examining the impact of managers' age, gender, tenure, and level of education on their choice between in-court and out-of-court restructuring. This adds to the debate on the key factors affecting the choice between in-court and out-of-court restructuring by highlighting manager tenure and level of education as new determinants. Finally, it sheds light on the effect of managers' demographic characteristics by showing that they do not affect the choice of the restructuring method. This is at least true in the context of distressed firms.

1. **UPPER ECHELONS THEORY AND FIRM RESTRUCTURING**

In this section, we first explain two restructuring approaches (in-court and out-of-court restructuring) for firms in financial distress (1.1). Further, we contrast the two approaches based on the upper echelons theory (1.2).

## 1.1. Two approaches for restructuring firms in financial distress

The choice of a restructuring strategy is a crucial decision and constitutes a defining event in the life of an organization (Bowman and Singh, 1993). Generally, when a firm faces financial difficulties, there are two approaches to resolving them. First, private restructuring consists of several decisions at the firm's financial or operational level (Dewaelheyns and Van Hulle, 2010; Gilson et al., 1990; Morrison, 2009; Jostarndt and Sautner, 2010). Second, in-court restructuring by initiating insolvency proceedings are equivalent to 'Procédure collective' in the French context of the present study. Its main objectives are to maintain the continuity of the business, preserve jobs, and repay creditors (Blazy et al., 2014).

Each restructuring method has its own unique features. In-court restructuring benefits from the legitimacy of the law in helping the firm. It obliges the creditor to wait for a common solution until the court rules on a plan for the distressed firm. At the same time, empirical evidence has shown that both the direct and indirect costs of insolvency proceedings are significant (Blazy et al., 2014). These costs are challenging to sustain, particularly for small businesses (Bergthaler, Kenneth, Yan, and Dermo, 2015). Direct insolvency proceedings costs include administrative and legal fees, and a firm's market value declines at the time of bankruptcy (Weiss, 1990). Indirect costs, which are generally more significant than direct costs, usually represent a wide range of opportunity losses and a loss of managers' ability to make their own decisions and exercise complete control over the firm's decisions (Hotchkiss et al., 2008). Finally, a firm's restructuring within insolvency proceedings may be affected by the efficiency of its process, which represents a significant indirect cost (Eckbo, Thorburn, and Wang, 2016).

While court restructuring incurs very high costs, researchers have shown that private restructuring allows for avoiding a large part of these costs (Hotchkiss et al., 2008). One of the primary benefits of private restructuring is confidentiality, which allows firms to negotiate all agreements with creditors to restructure the firm while continuing to operate. Out-of-court restructuring ensures that a firm's financial difficulties are handled confidentially, preserving creditors' trust and reputation among all stakeholders. Unlike private restructuring, insolvency proceedings are publicly exposed. Thus, initiating such a procedure may harm the firm's image, its relations with stakeholders, the reputation of its manager, and sometimes even the manager's professional career (Eckbo et al., 2016; Gilson et al., 1990; Hotchkiss et al., 2008;

Cusin, Gardès, and Maymo, 2022). Finally, private restructuring is generally faster than formal proceedings (Hotchkiss et al., 2008), given that court restructuring involves a multitude of complex interactions at multiple levels. Given these choices, a significant question among scholars has been to identify which factors favor in-court or out-of-court restructuring. They thus emphasized the importance of banking relationships (Peek and Rosengren, 2005), the nature of a firm's assets (Gilson et al., 1990), the firm's debt structure (Blazy et al., 2014), the dispersal of creditors (Franks and Sussman, 2005), and the asymmetry of information (Jostarndt and Sautner, 2010).

Surprisingly, prior studies have focused on firm characteristics; however, the characteristics of managers have received only limited attention, whereas the literature has elaborated on the impact of managers' characteristics on strategic decisions (Fisher and Chen, 2018; Benischke, Martin, and Glaser, 2019; Zemis and Demil, 2020; Khelil, 2016). Accordingly, in the context of distressed firms, we believe that managers' characteristics significantly influence their choice between in-court and out-of-court restructuring.

**1.2. Upper echelons theory: contrasting in-court and out-of-court restructuring**

Upper echelon theory is one of the most significant perspectives in strategic management (Hambrick and Mason, 1984). This lies in the idea that executives view their situations through their own highly personalized prisms. These individualized views of strategic situations are due to executives' differences in experience, values, and other human factors. Using upper echelon theory, researchers have investigated the effects of executive characteristics on firm strategy and performance. The logic behind this is that complex decisions are mostly the result of behavioral factors, including the organization's main actors' cognitive orientation, experiences, and personalities (Cyert and March, 1963).

Hambrick and Mason (1984) assert that managers' background characteristics and experiences influence the strategic decision-making process through their selective perception, limiting their field of vision and filtering internal and external information. Consequently, differences in managers' characteristics influence all stages of the strategic decision-making process: the identification of problems, selection and treatment of information, specification of actions, and proposal of solutions (Cyert and March, 1963).

Restructuring decisions may be subject to an executive's perception and interpretation of the level of risk associated with initiating a restructuring approach. Generally, out-of-court restructuring is riskier than in-court restructuring (Hotchkiss et al., 2008). The court protects the firm during insolvency proceedings. The manager is generally under less pressure, except in cases in which managers may be subject to professional sanctions or management suspensions. However, managers are under greater pressure during out-of-court restructuring because they are responsible for all firms' decisions. In other words, out-of-court restructuring is risky and likely for risk-takers. By contrast, in-court restructuring is less risky and more likely for risk-averse managers.

Moreover, these two approaches differ in the complexity of their processes. Specifically, out-of-court restructuring is more complex than in-court restructuring. Insolvency proceedings allow the firm's reorganization under the court's protection (Franks and Torous, 1989). During the insolvency procedure, judges assist the manager in developing a restructuring plan; they are not exclusively responsible for the firm's decisions. In contrast, the manager is the only person responsible for the firm's future and decisions within out-of-court restructuring, which requires a greater capacity to process information than in-court settings.

## 2. RESEARCH HYPOTHESES

This study focuses on managers' characteristics to understand the choice between in-court and out-of-court restructuring for firms in financial distress. Thus, in this section, we present our research hypothesis regarding managers' age (2.1), gender (2.2), position tenure (2.3), and educational level (2.4).

### 2.1. Manager Age

Hambrick and Mason (1984) argue that younger managers are generally associated with risk-taking and innovation in their strategic decisions, leading to more diverse and significant strategic decisions. As young managers do not have a reputation for high quality, they are subject to testing in the job market. Therefore, they are subject to stress and pressure from the firm and the job market, so they try to show their abilities by making riskier decisions. However, older managers are usually more conservative (Sundaram and Yermack, 2007). They become more risk-averse, capable, and ethical as they age (Malm, Adhikari, Krolikowski, and Nilesh, 2021).

Baker and Mueller (2002) show that younger managers often take more significant risks when investing than older managers. Prendergast and Stole (1996) developed a model of managerial signaling. They argued that young managers attempt to signal to the market that they possess great qualities and superior abilities by following high-risk strategies, including riskier and more aggressive investments. In particular, young managers exaggerate their personal beliefs and decision-making behavior to appear talented and stand out in the job market. On the other hand, older managers are reticent and conservative and, therefore, make less risky decisions.

Younger managers generally have less experience than their older peers and fewer opportunities to accumulate value and knowledge over their lives. Attracted by the prospect of professional career development and significant financial returns, younger managers may initiate aggressive strategic actions to generate personal and organizational value (Yim, 2013). Hambrick and Mason (1984) developed the following three ideas. First, with increasing age, leaders may come to a point in their lives when financial and job security are most important; therefore, they may engage in limited and less risky strategic actions. Second, older managers may be more attached to a firm's status and reputation. Finally, older managers may have lower mental endurance or be less able to capture new ideas and make risky decisions (Serfling, 2014). This may apply to the choice between in-court and out-of-court restructuring. Therefore, older managers are more likely to opt for in-court restructuring, which is less risky and provides court protection. However, younger managers are more likely to opt for private restructuring to demonstrate their ability to restructure the firm and gain more legitimacy.

*H1. Manager age is positively associated with in-court restructuring.*

## 2.2. Manager Gender

"*Maleness has become a synonym for insufficient attentiveness to risk.*" (Palvia, Vähämaa, and Vähämaa, 2015; Christopher Caldwell in Time, 2009, Vol. 174, No. 7:13).

Manager gender has been extensively studied in the decision-making literature and cognitive psychology and is related to conservatism, overconfidence, and risk acceptance (e.g., Thosuwanchot, 2021; Kaur and Singh, 2019; Croson and Gneezy, 2009). Previous studies have suggested that women generally think and behave differently than men. They have different preferences and understandings of risk (Habib and Hossain, 2013; Zalata, Tauringana, and Tingbani, 2018), which can significantly influence their decision-making process. Specifically, women are generally more risk-averse than men (Wei, 2007). Female managers are associated with more conservative, prudent, and less risky strategic decisions than male managers (Krishnan and Parsons, 2008; Barua, Davidson, Rama, and Thiruvadi, 2010). For instance, Huang and Kisgen (2013) find that firms with female managers are less likely to engage in debt issuance and acquisition strategies than firms with male managers. In addition, Walayet and Vieito (2013) show that the manager's gender matters in terms of firm performance, and when the manager is female, the firm's risk level is smaller than when the manager is male. More recently, Faccio, Marchica, and Mura. (2016) found that firms managed by women take fewer risks than those managed by men. More interestingly, to compare risk-taking in the same firms run by managers of different genders, they studied a sample of firms experiencing a transition from male to female managers or vice versa. They found that managers' transitions are associated with changes in corporate risk-taking. Specifically, transitions from male to female managers are associated with a reduction in corporate risk-taking.

In addition to risk-taking, Dunn (2012) noted that women are more likely than men to overestimate and fulfill their responsibilities as managers, which may explain why they are more inclined to use conservative and less risky management strategies (Zalata et al., 2018; Palvia et al., 2015). However, men are more motivated by increasing economic benefits and are likelier to break the rules to achieve great success and high performance. Similarly, Ho et al. (2015) found that female managers are more ethical in decision-making. They are more compliant with regulations (Butz and Lewis, 1996), which may be consistent with dealing with financial distress before the court and not procrastinating the decision to do so.

Based on the previous discussion, female managers are more risk-averse, conservative, and rule-following than male managers are. This behavior may be reflected in their decision-making, as in the choice of a restructuring strategy. Therefore, we expect distressed firms with female managers to be more likely to opt for in-court restructuring (i.e., less risky than out-of-court restructuring) by seeking court protection and avoiding management errors. However, firms with male managers are more likely to choose out-of-court restructuring by privately reorganizing the firm.

*H2. Firms with female managers will be more likely to opt for in-court restructuring. However, firms with male managers will be more likely to opt for out-of-court restructuring.*

## 2.3. Manager Position Tenure

Manager tenure is one of the most studied characteristics in upper echelons theory research (Finkelstein et al., 2009). Several studies have examined the effect of manager tenure on several firms' strategic decisions by using tenure level as a proxy for persistence, status-quo engagement, and rigidity (Finkelstein and Hambrick, 1996). For instance, the effect of manager tenure on mergers and acquisitions (Zhou, Shantanu, Pengcheng, 2020), corporate social responsibility (Chen, Zhou, and Zhu, 2019), commitment to the status quo (Musteen, Barker, and Baeten, 2006), strategic change (Zhang and Rajagopalan, 2010), firm value (Brookman and Thistle, 2009), innovation (Li and Yang, 2019), managers' cognitive complexity (Graf-Vlachy, Bundy, and Hambrick, 2020), and risk-taking (Boling, Pieper, and Covin, 2016).

Research has shown that expertise is a continuum, and managers progressively acquire more expertise—the ability to process complex information—as they accumulate experience (Graf-Vlachy et al., 2020). Managers who face difficult tasks gain more experience and expertise (Ericsson, 2005). At least in their new roles, they may be considered novices at the beginning of their tenure. Even if they have distinguished themselves from previous positions, new managers face problems and tasks that are largely unfamiliar to them (Hambrick and Fukutomi 1991). In addition, even if they were previously familiar with their field of activity, the new environment was different and required time to adapt. In particular, the early years of a manager's tenure tend to be marked by limited firm knowledge, making it more difficult to process complex information and make risky strategic decisions. This sometimes results in new managers being unable to handle or perhaps even recognize the complexity they face (Graf-Vlachy et al., 2020). As managers advance in their tenure, their expertise increases, and they have both the flexibility and knowledge to process more complex information (Ericsson, 2005). In line with this, several studies argue that managers with a high tenure may make risky strategic decisions because of their familiarity with their company- or industry-specific business experience (Carpenter et al., 2003).

As tenure increases, managers become more comfortable with decision-making and gain organizational legitimacy (Abebe, 2009). Over time, managers who hold their positions accumulate experience, become familiar with the firm's culture, and develop more relationships with all stakeholders (Finkelstein and Hambrick, 1996), making it easier to negotiate and restructure the firm privately. The legitimacy earned by an increase in tenure strengthens their impact on corporate decisions and protects them from the pressures of economic performance (Boeker, 1989). In addition, gaining employee trust, coupled with the accumulation of experience within the firm, gives managers a more significant position and power, allowing them to adopt riskier strategies (Hung and Tsai, 2020). Furthermore, higher tenure is associated with greater autonomy (Miller, 1991), which gives managers confidence and increased opportunities to pursue strategic options that involve a higher level of risk.

Furthermore, initiating insolvency proceedings can result in negative reputational consequences for the firm and the manager. Thus, potentially damaging their relationships with stakeholders and exposing them to stigmatization (Sutton and Callahan, 1987; Cardon, Stevens, and Potter, 2011). Managers with higher tenure have likely built up a reputation and a career within the company and are more invested in its success. Initiating a formal restructuring process can potentially damage the manager's reputation and career prospects, given the stigma associated with it, which is a significant risk for someone who has worked with the company for a long time and has much to lose compared to managers with lower tenure. On the other hand, a newly recruited manager may be more willing to take on this risk since they have not yet had time to establish a reputation within the company. Additionally, a higher-tenure manager may have more knowledge, skills, and contacts within the company, making it easier to restructure privately without resorting to a formal process.

Based on these arguments, it turns out that newly selected managers are likely to prefer options that involve less information processing, thus favoring restructuring before the court. However, managers are more able to process information, deal with complex situations, and gain more expertise as their position tenure increases (Graf et al., 2020). Consequently, they absorb the complexity of their situation and gain greater legitimacy and power over the firm, which may make out-of-court restructuring preferable. In addition, as managers advance in their careers, they become more concerned with their legacy and status quo (Matta and Beamish, 2008). Thus, given that in-court restructuring is publicly exposed, managers with a higher tenure are more likely to opt for private restructuring to avoid a negative impact on their reputation and the firm's image (Eckbo et al., 2016).

Based on this discussion, higher-tenured managers are likely to choose out-of-court restructuring. However, those with less tenure are more likely to choose in-court restructuring.

*H3. Manager tenure is negatively associated with in-court restructuring.*

## 2.4. Manager Educational Level

Previous studies have argued that formal education is the primary basis of cognitive orientation (Hambrick and Mason, 1984). Educational level has been associated in the upper echelons literature with tolerance for ambiguity, capacity for information processing, and the ability to identify and evaluate multiple alternatives. This may explain why differences in managers' education levels strongly influence their decision-making processes and, consequently, the firm's strategic choices (Finkelstein et al., 2009; Lewis, Andriopoulos, and Smith, 2014). For instance, Malmendier and Tate (2005) found that managers' educational level significantly impacts a firm's investment strategy. Barker and Mueller (2002) provide evidence of the positive effect of managers' educational levels on market innovation strategies.

Dealing with financial difficulties requires superior knowledge and understanding of all business details; thus, processing complex information from multiple sources is important in such a situation. Out-of-court restructuring usually involves more in-depth information analysis among the various stakeholders in the negotiation process, which requires more precise decision-making. However, within in-court restructuring, the court supports the manager to some extent and is no longer the only person responsible for the decisions taken. Therefore, it does not necessarily require a high level of information processing capacity compared to out-of-court restructuring, where the manager is exclusively accountable for the decisions and, thus, for the firm's future.

For several reasons, we suppose a negative relationship between in-court restructuring and managers' educational level. First, according to Wally and Baum (1994), increasing a manager's level of formal education enhances their cognitive capacity, which helps them acquire and process more complex information and make riskier decisions. Second, formal education helps managers accumulate valuable knowledge and develop curiosity and openness toward new challenges. Third, formal education can also give managers the skills to understand complex situations and manipulate information when a firm faces financial difficulties. Finally, as Thomas, Litscert, and Ramaswamy (1991) have argued, highly educated managers are often more receptive to new challenges, which may increase their motivation to adopt complex and risky business strategies. Given the above reasons and the difficulty of the challenge when a firm is in financial distress, we suppose that managers with a higher level of formal education will be more likely to opt for out-of-court restructuring.

*H4. Managers' educational level is negatively associated with in-court restructuring.*

## 3. METHODOLOGY

The following section presents the methodology used in this study. To do so, we describe the composition of our sample, data collection, variable measurement, and the analyses performed to test the different hypotheses.

### 3.1. Sample and Data Collection

Our sample consisted of 342 top managers from French SMEs who qualified as financially distressed in 2019. A four-step selection process was used to identify the sample. First, we extracted the target population from the DIANE database, which contains the main financial information about French companies. The extracted data included firm-specific information, financial statements, and contact details. Second, we targeted unlisted financially distressed firms that could potentially restructure their situations. We used the definition of Álvarez Román, Garcia-Posada, and Mayordomo (2022) to select financially distressed firms. They define a company as financially distressed if it is at least 5 years old and has both an interest coverage ratio lower than one (the ratio of a company's EBITDA to its interest expense) and negative equity for at least 3 consecutive years. Based on the more reliable financial data of the "pre-Covid pandemic financial situation of firms" for 2019, we retain only firms that meet those conditions. Third, we select only SMEs (employees <=250), according to the definition of SMEs, based on the recommendations of the European Commission (2003). Finally, only SMEs with valid email addresses and manager contact information were retained. The constructed questionnaire was pilot tested, pretested, and administered directly to managers.

We contacted the managers by asking them to answer the survey. The invitation contained a cover letter explaining that the study was supported by the Commercial Court of Saint Etienne and CIP Loire Sud (Centre d'Information sur la Prévention des Difficulties des Entreprises). This organization promotes the measures provided by French law to prevent or deal with business difficulties. We recorded a final sample of 342 responses that provided all the required data. Finally, we obtain a very satisfactory final sample corresponding to a response rate of 12%, consistent with previous similar studies of managers (Herrmann and Nadkarni, 2014), and is appropriate for research of this nature in the context of declining executive response rates (Garcés-Galdeano et al., 2017).

**3.2. Measures**

This section presents the study's variables: dependent variable (3.2.1), independent variables (3.2.2), and finally, the control variables (3.2.3).

*3.2.1. Dependent variable*

The dependent variable of this study is binary, coded 1 if the manager is more likely to opt for in-court restructuring and 0 if the manager is more likely to opt for out-of-court restructuring. To measure our dependent variable, we asked managers to indicate the measures they intended to implement to address their difficulties. We propose 11 restructuring measures, with the option of choosing more than one measure. Four French bankruptcy-specific insolvency proceedings (Mandat Ad Hoc, Conciliation, Sauvegarde, Redressement Judiciaire) and seven private restructuring measures (e.g., employee downsizing and searching for new financing). Since private restructuring measures are also used in formal restructuring, it is evident that all managers have chosen at least some private restructuring measures. Therefore, we coded the dependent variable as 1 if the manager chooses at least one of the four insolvency proceedings and 0 if the manager chooses none.

To strengthen the robustness of our results and deepen our understanding of the phenomenon under investigation, we conducted supplementary tests using an alternative measure of our dependent variable. In contrast to our initial binary measure, we employed a continuous measure by asking CEOs to rate their intention to initiate insolvency proceedings on a four-level Likert scale ranging from "low intention" to "high intention." We proposed four bankruptcy procedures specific to French bankruptcy law ("Mandat Ad Hoc," "Conciliation," "Sauvegarde," and "Redressement Judiciaire"), allowing respondents to choose more than one answer. The resulting measure ranged from 4 to 16, with higher scores indicating stronger intentions to choose an in-court restructuring. Our sample yielded an average in-court restructuring score of approximately 5.961. To assess the dimensionality of our dependent variable measure, we conducted a confirmatory factor analysis (CFA). The CFA results confirmed the unidimensionality of the measure of in-court restructuring, as indicated by the comparative fit index (CFI = 0.93) and the root mean square error of approximation (RMSEA = 0.03). Moreover, we assessed the reliability of the measure using Cronbach's alpha and found it to be satisfactory (alpha = 0.87). We report the results of these robustness tests in Table 4 and confirm that they did not substantially modify our original results. These tests provide further evidence of our results' validity and robustness.

### *3.2.2. Independent variables*

Manager age is measured as the natural logarithm of the number of years the manager was alive at the time of data collection. Manager gender is a dummy variable that takes 1 if the manager is a woman and zero if the manager is a man. Manager tenure is measured as the natural logarithm of the number of years the individual has held the manager's position in their current firm (Miller, 1991). Formal manager education is based on their level of schooling. We used a six-point scale: 1=primary education (École secondaire), 2= middle school (collège), 3= high school (lycée: Baccalauréat), 4 = higher education (LMD Framework: Bac+2 / Bac+3: niveau BTS, Licence), 5= higher education (LMD framework: Master, Ingénieurs), and 6= higher education (LMD framework: Doctorat).

### *3.2.3. Control Variables*

We controlled for several factors that may impact restructuring decisions. At the firm level, we first controlled for firm size, which may affect the decision to restructure before the court (Claessens and Klapper, 2005). Small firms may avoid formal proceedings to prevent high costs. The number of employees approximates the variable size. Second, we control the firm's industry, given that the level of failure changes across industries (Morrow, Johnson, and Busenitz, 2004). We used 10 dummies corresponding to different industries: agriculture and fishing, transport and logistics, hotels and restaurants, construction, trade, industry, information and communication, business services, individual services, and others. Moreover, we controlled for financial variables that may impact firms' restructuring decisions. Given that financially distressed firms are expected to reduce their liabilities to suppliers, investments, and loans to regain their economic strength, we controlled for the level of debt (debt over total assets). We also controlled for firm turnover. Further, we controlled the level of cash (cash and equivalents/assets) (Fisher et al., 2022). Finally, we controlled for firm age as a proxy for a firm's reputation. This is expected to negatively influence the probability of choosing an in-court restructuring.

At the manager level, we first control for the power and influence of the manager by controlling whether the manager is the founder using a dummy variable coded 1 if the manager is the founder and 0 otherwise. Second, we controlled for the manager's functional experience (Barker and Mueller, 2002). According to UET, managers' prior functional experiences shape their strategic choices by influencing how they interpret information based on their expertise. Thus, we controlled the manager's functional experience by using a dummy variable coded 1 if the manager has more "output" experience (sales/marketing, product R&D, and entrepreneurship) and coded 0 if they have more "throughput" experience (production/operations, finance, accounting, data treatment/information systems, and process R&D). We further assessed the manager's experience in managing financial difficulties by asking them, "Have you ever managed a financially distressed firm? "If yes, how many times?". Additionally, we controlled their experience with insolvency proceedings by asking them, "Have you ever initiated insolvency proceedings for a firm?".

## 4. RESULTS

This section presents the descriptive analysis of our data (5.1) and the hypothesis testing (5.2).

**4.1. Descriptive analysis**

**Table 1. Descriptive statistics of variables**

| Variable | Mean | Std. Dev. | Min | Max |
|---|---|---|---|---|
| *Firm-level* | | | | |
| In-court restructuring | 0.066 | 0.249 | 0 | 1 |
| Firm age | 27.983 | 23.196 | 5 | 192 |
| Firm size (number of employees) | 34.614 | 51.483 | 0 | 250 |
| Turnover | 6653421.2 | 14557828 | 0 | 133000000 |
| Financial debts/Assets | 0.254 | 4.638 | 0 | 559.8 |
| Cash and equivalent /Assets | 0.015 | 0.226 | 0 | 36.11 |

| **Firm industry** | | | **Freq.** | **Percent** |
|---|---|---|---|---|
| Industry | | | 81 | 23,68 % |
| Construction | | | 41 | 11,99 % |
| Trade | | | 46 | 13,45 % |
| Agriculture and fishing | | | 6 | 1,75 % |
| Information and communication | | | 19 | 5,56 % |
| Hotels and restaurants | | | 7 | 2,05 % |
| Services to business | | | 73 | 21,35 % |
| Services to individuals | | | 13 | 3,80 % |
| Transport and logistics | | | 14 | 4,09 % |
| Other services | | | 42 | 12,28 % |

*Manager-level*

| | Mean | Std. Dev. | Min | Max |
|---|---|---|---|---|
| Manager age | 51.449 | 9.562 | 23 | 81 |
| Manager gender | 0.215 | 0.411 | 0 | 1 |
| Manager tenure | 15.289 | 10.054 | 0 | 50 |
| Manager output experience | 0.556 | 0.497 | 0 | 1 |
| Experience with financial distress | 1.358 | 2.08 | 0 | 15 |
| Experience with insolvency proceedings | 0.234 | 0.523 | 0 | 3 |
| Manager founder | 0.595 | 0.492 | 0 | 1 |

| **Manager educational level** | | | **Freq.** | **Percent** |
|---|---|---|---|---|
| Primary education (École secondaire) | | | 12 | 3.51 % |
| Middle school (collège) | | | 11 | 3.22 % |
| High school (lycée: Baccalauréat) | | | 27 | 7.89 % |
| Higher education (LMD Framework: Bac+2 / Bac+3) | | | 95 | 27.78 % |
| Higher education (LMD framework: Master, Ingénieurs) | | | 180 | 52.63 % |
| Higher education (LMD framework: Doctorat) | | | 17 | 4.97 % |

Table 1 provides the descriptive statistics for the variables in this study. The mean value of in-court restructuring is 0.066, suggesting that 6.6% of managers prefer to restructure before the court (in-court), and 93.4% of managers prefer to restructure privately (out-of-court). The average age of the managers in this study was 51.45 years, and 21.5% were women. These statistics are similar to those of previous studies investigating manager characteristics (e.g., Ho, Tam, and Zhang, 2015). In our sample, 55.6% of the managers have more "output" functional experience, and 44.4% have more "throughput" functional experience. In addition, we report that the average mean manager tenure was 15.29 years.

Table 2 presents the correlation matrix between the study's variables. We can observe that the variables do not represent high correlation coefficients. We used variance inflation factors (VIFs) to control for a possible multicollinearity problem. The correlations and the VIFs observed between the variables are satisfactory and do not present any particular concern for the multivariate analysis.

**Table 2. Correlations among the research variables**

| Variables | (1) | (2) | (3) | (4) | (5) | (6) | (7) | (8) | (9) | (10) | (11) | (12) | (13) | (14) | **VIF** |
|---|---|---|---|---|---|---|---|---|---|---|---|---|---|---|---|
| (1) In-court restructuring | 1.000 | | | | | | | | | | | | | | - |
| (2) Manager age | -0.146*** | 1.000 | | | | | | | | | | | | | 1.43 |
| (3) Manager gender | 0.104** | -0.143*** | 1.000 | | | | | | | | | | | | 1.10 |
| (4) Manager tenure | -0.194*** | 0.524*** | -0.103* | 1.000 | | | | | | | | | | | 1.60 |
| (5) Manager educational level | -0.135** | -0.101* | -0.105** | -0.206*** | 1.000 | | | | | | | | | | 1.11 |
| (6) Manager output experience | 0.014 | 0.008 | -0.059 | 0.062 | -0.129** | 1.000 | | | | | | | | | 1.05 |
| (7) Manager experience with financial distress | -0.041 | 0.127** | -0.132** | 0.118** | -0.023 | -0.006 | 1.000 | | | | | | | | 1.17 |
| (8) Manager experience with insolvency proceedings | 0.093* | 0.179*** | -0.055 | 0.085* | 0.014 | -0.088* | 0.331*** | 1.000 | | | | | | | 1.21 |
| (9) Manager founder | 0.039 | 0.167*** | -0.019 | 0.303*** | -0.183*** | 0.099* | 0.131** | -0.006 | 1.000 | | | | | | 1.52 |
| (10) Firm age | -0.065 | 0.200*** | -0.057 | 0.164*** | 0.048 | -0.080 | -0.006 | 0.105** | -0.372*** | 1.000 | | | | | 1.36 |
| (11) Firm size (number of employees) | -0.082 | 0.102* | -0.128** | 0.010 | 0.102* | -0.026 | 0.069 | 0.183*** | -0.295*** | 0.234*** | 1.000 | | | | 1.52 |
| (12) Turnover | -0.129** | 0.131** | -0.078 | 0.192*** | -0.012 | -0.059 | 0.073 | 0.095* | -0.136*** | 0.198*** | 0.335*** | 1.000 | | | 1.63 |
| (13) Financial debts/Assets | -0.109** | 0.116** | -0.077 | 0.141*** | 0.059 | -0.071 | 0.022 | 0.038 | -0.090* | 0.157*** | 0.218*** | 0.484*** | 1.000 | | 2.02 |
| (14) Cash and equivalent /Assets | 0.069 | -0.089* | -0.019 | -0.142*** | -0.054 | 0.051 | 0.013 | 0.062 | 0.047 | -0.111** | 0.059 | -0.144*** | -0.379*** | 1.000 | 1.30 |
| (15) Firm industry | 0.020 | -0.023 | 0.104** | -0.110** | 0.127** | -0.018 | -0.015 | 0.032 | 0.078 | -0.136*** | -0.035 | -0.083 | -0.041 | -0.031 | 1.08 |

*\*\*\* p<0.01, \*\* p<0.05, \* p<0.1*

**Table 3. Marginal effects (dy/dx) of a binary logistic regression**

|  | Model 1 | Model 2 | Model 3 | Model 4 | Model 5 |
|---|---|---|---|---|---|
| Manager age | -.112* |  |  |  | -.036 |
|  | (.061) |  |  |  | (.060) |
| Manager gender |  | .049 |  |  | .036 |
|  |  | (.027) |  |  | (.026) |
| Manager tenure |  |  | -.041*** |  | -.045*** |
|  |  |  | (.015) |  | (.016) |
| Manager educational level |  |  |  | -.023** | -.027*** |
|  |  |  |  | (.011) | (.011) |
| Manager output experience | -.016 | -.013 | -.012 | -.024 | -.015 |
|  | (.025) | (.025) | (.025) | (.025) | (.024) |
| Manager financial distress experience | -.006 | -.005 | -.007 | -.007 | -.004 |
|  | (.008) | (.008) | (.009) | (.008) | (.008) |
| Manager insolvency proceedings experience | .043* | .037* | .042* | .040* | .044** |
|  | (.023) | (.022) | (.023) | (.022) | (.022) |
| Manager founder | .016 | .004 | .042 | -.009 | .042 |
|  | (.028) | (.027) | (.031) | (.027) | (.032) |
| Firm age | .000 | -.000 | .000 | -.000 | .001 |
|  | (.001) | (.001) | (.001) | (.001) | (.001) |
| Firm size (number of employees) | .002 | .002 | .000 | .001 | .005 |
|  | (.008) | (.009) | (.008) | (.008) | (.008) |
| Turnover t-1 | -.009** | -.010*** | -.009** | -.010*** | -.008** |
|  | (.004) | (.004) | (.004) | (.004) | (.004) |
| Financial debts/ Assets t-1 | .003 | .003 | .003 | .002 | .002 |
|  | (.007) | (.007) | (.007) | (.007) | (.006) |
| Cash and equivalent / Assets t-1 | .005 | .008 | .002 | .005 | -.002 |
|  | (.014) | (.014) | (.014) | (.013) | (.015) |
| Firm Industry | -.005 | -.007 | -.006 | -.005 | -.006 |
|  | (.005) | (.006) | (.006) | (.005) | (.005) |
|  |  |  |  |  |  |
| R-squared | .127 | .126 | .166 | .136 | .236 |
| Maximum VIF | 2.00 | 2.00 | 2.00 | 2.01 | 2.02 |
| Correctly Predicted | 89.26 % | 89.26 % | 88.98 % | 89.26 % | 89.26 % |
| Observations | 342 | 342 | 342 | 342 | 342 |

Notes: The standard errors are shown in parentheses. The coefficients are estimated using a logistic regression model. The maximum VIF represents the highest value of the Variance Inflation Factor among the independent variables. The percentage of cases correctly predicted by the regression is reported as "Correctly Predicted." Asterisks indicate statistical significance: *p<.1, **p<.05, ***p<.01*.

|  | Model 6 | Model 7 | Model 8 | Model 9 | Model 10 |
|---|---|---|---|---|---|

**Table 4. Results of Ordinary Least Squares (OLS) regression**

| | | | | | |
|---|---|---|---|---|---|
| Manager age | -.634* | | | | -.055 |
| | (.184) | | | | (.045) |
| Manager gender | | .031 | | | -.022 |
| | | (.085) | | | (.085) |
| Manager tenure | | | -.093** | | -.564*** |
| | | | (.041) | | (.203) |
| Manager educational level | | | | -.053** | -.069** |
| | | | | (.035) | (.035) |
| Manager output experience | .068 | .066 | .076 | .055 | .06 |
| | (.067) | (.069) | (.068) | (.069) | (.068) |
| Manager financial distress experience | -.005 | -.006 | -.006 | -.007 | -.006 |
| | (.017) | (.017) | (.017) | (.017) | (.017) |
| Manager insolvency proceedings experience | .05* | .042 | .039 | .044 | .051* |
| | (.03) | (.03) | (.03) | (.03) | (.03) |
| Manager founder | .075 | .004 | .077 | -.016 | .086 |
| | (.078) | (.077) | (.083) | (.077) | (.082) |
| Firm age | -.001 | -.003* | -.002 | -.003* | -.001 |
| | (.002) | (.002) | (.002) | (.002) | (.002) |
| Firm size (number of employees) | .086 | .061 | .07 | .062 | .089 |
| | (.069) | (.069) | (.069) | (.069) | (.068) |
| Turnover t-1 | -.083*** | -.087*** | -.082*** | -.086*** | -.08*** |
| | (.017) | (.017) | (.017) | (.017) | (.017) |
| Financial debts/ Assets t-1 | .001 | .003 | .004 | 0 | -.001 |
| | (.022) | (.022) | (.022) | (.022) | (.022) |
| Cash and equivalent / Assets t-1 | .02 | .033 | .019 | .027 | .006 |
| | (.045) | (.046) | (.046) | (.046) | (.046) |
| Firm Industry | -.017 | -.02 | -.021 | -.017 | -.015 |
| | (.014) | (.015) | (.015) | (.015) | (.015) |
| _Cons | 4.827* | 2.484* | 2.589* | 2.76* | 4.977* |
| | (.732) | (.286) | (.284) | (.33) | (.809) |
| | | | | | |
| R-squared | .159 | .129 | .142 | .135 | .172 |
| Maximum VIF | 2.00 | 2.00 | 2.00 | 2.01 | 2.02 |
| Observations | 342 | 342 | 342 | 342 | 342 |

Notes: The standard errors are shown in parentheses. The coefficients are estimated using a linear regression model. The maximum VIF represents the highest value of the Variance Inflation Factor among the independent variables. Asterisks indicate statistical significance: *$p<.1$, ** $p<.05$, *** $p<.01$.

### 4.2. Hypothesis Testing

In light of the binary dependent variable in our study, we employed logistic regression to examine the influence of manager characteristics on the choice between in-court and out-of-court restructuring. The logistic regression results, presented in Table 3, revealed interesting findings. In particular, we conducted a marginal effects analysis to gain deeper insights into the relationship between managers' characteristics and their likelihood of initiating insolvency proceedings.

Firstly, it is noteworthy that managers' sociodemographic variables, including age ($dy/dx = -0.036$, $p > 0.1$, Model 5) and gender ($dy/dx = 0.036$, $p > 0.1$, Model 5), do not appear to affect the decision to choose between in-court and out-of-court restructuring. Thus, hypotheses H1 and H2 were not supported.

Secondly, we observed a negative effect of manager tenure on the probability of choosing in-court restructuring ($dy/dx = -0.045$, $p < 0.01$, Model 5). The results indicate that with every one-unit increase in managerial tenure, there is a significant reduction by 4.5% in the probability of choosing insolvency proceedings, with all other variables remaining at their means. This suggests that managers serving longer are less likely to initiate legal proceedings and prefer resolving matters outside of court. In other words, managers with higher tenures are less inclined towards in-court restructuring and more likely to favor out-of-court restructuring. Thus, hypothesis H3 is supported.

Finally, our analysis sheds light on the impact of educational level on initiating insolvency proceedings. We found that a one-unit increase in the manager's education level led to a 2.7% decrease in the likelihood of selecting formal court-based procedures over out-of-court mechanisms ($dy/dx = -0.027$, $p < 0.01$, Model 5). Consequently, our results suggest that managers with higher educational qualifications are less inclined towards in-court restructuring methods than their counterparts with lower qualifications. Thus, hypothesis H4 is supported.

To ensure the robustness of our findings, we performed an additional analysis using a continuous dependent variable, as presented in Table 4. The results obtained through the ordinary least squares (OLS) approach reaffirmed the conclusions drawn from our previous analysis. Furthermore, we conducted a comparative analysis by examining the results using a subsample specific to the Rhône-Alpes region. The Rhône-Alpes region holds the second position, after Ile de France, regarding bankruptcy registrations and economic significance, making it an ideal subsample for our study. The analysis results, presented in Appendix A.1, validate and reinforce our main findings. We employed the variance inflation factor (VIF) for each predictor to address the possibility of multicollinearity among independent variables.

Severe multicollinearity would be indicated by a VIF value exceeding 10, raising concerns about the econometric accuracy. However, upon analyzing our data, we found none of these issues were present, as all recorded VIF values were below 2.02 (Table 3 and Table 4).

## 5. DISCUSSION

After discussing the results of our study (5.1), its theoretical contributions, and practical implications (5.2), we present the limitations and future research direction of this study (5.3).

### 5.1. Results discussion

This study addresses an important gap in the empirical literature by examining the impact of manager characteristics on the choice between in-court and out-of-court restructuring for firms in financial distress. Overall, our empirical analysis provides interesting insights. First, this study shows that, with increasing tenure, managers are more likely to prefer out-of-court restructuring to in-court restructuring. This result is in line with the theoretical arguments. This could mean that the higher ability for information processing (Graf-Vlachy et al., 2020), dealing with complex situations, experience, and greater legitimacy in their position probably made managers with a higher tenure more inclined to prefer private (out-of-court) restructuring.

Furthermore, the preference of managers with longer tenure for private restructuring over insolvency proceedings could be partly explained by the stigma associated with insolvency proceedings. The initiation of insolvency proceedings can result in negative reputational consequences for the firm and the manager. Thus, potentially damaging their relationships with stakeholders and exposing them to stigmatization (Sutton and Callahan, 1987; Cardon, Stevens, and Potter, 2011). These detrimental effects can exacerbate the stigma attached to the firm and its management, leading to long-term implications such as limited access to credit, difficulties in attracting new investors, and a negative impact on the manager's professional career (Sutton and Callahan, 1987). Therefore, managers with longer tenure, who have invested significant time and effort in building their reputation and the firm's reputation, maybe more hesitant to opt for insolvency proceedings and prefer private restructuring to avoid the associated stigma.

Second, this study found a significant negative relationship between managers' educational level and the probability of choosing in-court restructuring. We found that the higher the manager's educational level, the lower the likelihood of in-court restructuring. This finding is consistent with the overall argument in the upper echelons theory. Accordingly, this could mean managers with higher educational levels have cognitive orientations and the ability to analyze complex information and issues (Wally and Baum, 1994), which helps them accept the risks associated with private restructuring. Out-of-court restructuring typically entails a more in-depth analysis of information in the restructuring process and more precise decision-making.

Finally, our results show that managers' demographic variables, including age and gender, do not significantly affect their in-court and out-of-court restructuring preferences. This result is contrary to the theoretical arguments. In the following discussion, we provide insights to explain this result. We built our first hypothesis on the argument that manager age is positively associated with risk aversion (Serfling, 2014; Ashton and Lee, 2016). However, some previous studies have reported that older managers are likelier to make risky decisions (McClelland, Barker, and Oh, 2012). Similarly, Holmstrom (1999) reported that young managers are more risk-averse, face significant professional concerns, and worry more about their reputations, leading to excessive conservatism in their strategic decisions. Therefore, we conclude that the influence of age on risk aversion is not in one direction and changes from one strategic context to another, which explains the contradictory results of previous studies and our outcome.

Although we did not find a significant relationship between manager gender and the choice between in-court and out-of-court restructuring, previous studies have identified this variable (i.e., manager gender) as important in shaping corporate decisions and outcomes (e.g., Weber and Zulehner, 2010; Berger, Kick, and Schaeck, 2014). Our result is consistent with a recent study by Vathunyoo, Angelica, and Jens (2016) on a large sample of US firms drawn over 15 years, which provides evidence that gender diversity does not affect firm risk-taking after controlling for the endogeneity of gender selection.

Moreover, the research by James (2016) sheds light on the strategic use of legal restructuring procedures by managers to enhance their firm's financial value and competitiveness. As demonstrated in our study, this may explain why certain demographic variables, such as age and gender, have insignificant effects on managers' restructuring preferences. Instead, cognitive variables, specifically educational level and tenure, significantly impact a manager's ability to comprehend and initiate these procedures. A manager with a higher educational level is likely to possess a more nuanced understanding of the legal and financial implications of these procedures and apply them more proficiently to improve the firm's financial standing. Therefore, cognitive variables, such as educational level and tenure, exert a more substantive influence on a manager's restructuring decisions than demographic variables.

Thus, our results support the fact that despite the extensive use of demographic factors, these factors may not fully capture the cognitive variables of interest (Barker and Mueller, 2002). Therefore, our conclusion is straightforward: the gender of the executive or the proportion of women on a board of directors will not affect the preferences and the choice between in-court and out-of-court restructuring. This conclusion is particularly relevant for the boards of directors.

## 5.2. Theoretical contributions and practical implications

This study contributes to and expands the literature in several ways. First, it contributes to the upper echelons theory (Hambrick and Mason, 1984) by investigating the impact of manager characteristics on restructuring decisions. Second, this study contributes to the literature on bankruptcy and restructuring. It enriches our understanding of the determinants of the restructuring decision "in-court and out-of-court dichotomy" by moving its focus from firm-specific characteristics to the manager's characteristics. It provides evidence that managers generally resist court intervention. At the same time, their intentions vary neatly according to their characteristics, such as educational level and position tenure, which should be considered. Finally, we show that manager age and gender do not affect the restructuring decision, unlike other studies (Walayet and Vieito, 2013; Fuming Jiang et al., 2018; Liu et al., 2018) that have shown a significant effect of demographic variables on other strategic decisions.

This study also has several practical implications for managers, board directors, auditors, and legislators. First, this study's findings show managers' characteristics may affect restructuring decisions. Accordingly, managers must understand that their characteristics may affect their restructuring choices. Therefore, our results provide an opportunity to rethink managers' restructuring choices and keep the firm's financial situation in mind to increase its chances of a successful turnaround. Second, legislators may consider the characteristics of the executive when designing insolvency proceedings. Third, accountants and auditors must be more careful with different manager profiles in their audit and accounting missions. Some managers with high education and tenure levels may prefer to restructure privately and postpone the decision to file for insolvency proceedings. This can sometimes lead to severe failure and decrease the chances of a successful turnaround. Finally, boards of directors are strongly encouraged to consider the traits and characteristics of executives when choosing a top manager, at least when making the most critical strategic decisions.

**5.3. Limitations and avenues for future research**

Despite these theoretical and practical implications, this study had some limitations that should be considered and addressed in future research. Firstly, our sample was restricted to relatively small- and medium-sized French firms, limiting our results' generalizability. Future research examining the relationships investigated in this study with larger/listed firms should provide further insights.

Secondly, it can be argued that managers are likely to be influenced by their psychological attributes and personality traits in their decisions. We believe that future research investigating how manager personality and psychological attributes (Nadkarni and Herrmann, 2010) help understand the restructuring decision would undoubtedly improve our understanding of the factors influencing restructuring choices. For instance, we can study personality traits using the Big Five factors (Nadkarni and Herrmann, 2010; Caliendo, Fossen, and Kritikos, 2014) or psychological attributes, such as narcissism (Cragun, Olsen, and Wright, 2020; Lin and Fang, 2020) and optimism (Hung and Tsai, 2020), two of the most studied psychological attributes in the strategic decision literature.

Thirdly, future research could overcome the limitations of this study by expanding on our findings and exploring the factors that affect managers' restructuring preferences in more detail. For instance, future studies could collect data on both firm and manager characteristics to gain a more comprehensive understanding of the decision-making process. This could lead to a more nuanced exploration of the factors influencing managers' restructuring preferences, identifying additional variables not accounted for in our research. Furthermore, a comprehensive exploration of the strategic use of legal restructuring procedures, considering both cognitive and demographic variables and firm characteristics, could lead to more effective strategies for managing financial distress and promoting the long-term financial health of firms.

Finally, it would be interesting to investigate the real decisions of firms in comparison to managers' intentions using a longitudinal approach, as our analysis was based on managers' intentions to proceed with in-court or out-of-court restructuring to address their financial difficulties.

## CONCLUSION

In conclusion, our study sheds light on the impact of manager characteristics on the choice between in-court and out-of-court restructuring, building on the upper echelons theory. The results demonstrate that specific manager characteristics play a significant role in the decision-making process regarding restructuring decisions. Specifically, we find that managers with longer tenure and managers with higher education are less likely to choose in-court restructuring. On the other hand, we find no significant impact of age or gender on restructuring decisions. This implies that these demographic factors may not play a significant role in the decision-making process for French SME managers when considering restructuring options. However, further research is needed to investigate the impact of these factors in other contexts and settings. These results contribute to the literature on upper echelons theory and provide practical implications for policymakers and practitioners involved in the restructuring process. Future research could explore the impact of other manager characteristics and firm-level factors on restructuring decisions.

## ACKNOWLEDGMENTS

I am very grateful to the two anonymous reviewers for their substantial contribution to improving this article. Additionally, I sincerely thank Marc Frechet, Ivana Vitanova, and Frederic Perdreau for their insightful discussions and valuable suggestions.

**DECLARATION OF COMPETING INTEREST**

The author declares that no known competing financial interests or personal relationships could have appeared to influence the work reported in this paper.

**FUNDING**

This research did not receive specific grants from public, commercial, or not-for-profit funding agencies.

# Appendix

**Table A.1. Robustness test (OLS) results for the subsample of the Rhône-Alpes region**

|  | Model 11 | Model 12 | Model 13 | Model 14 | Model 15 |
| --- | --- | --- | --- | --- | --- |
| Manager age | -.051* |  |  |  | -.068 |
|  | (.035) |  |  |  | (.035) |
| Manager gender |  | .026 |  |  | -.027 |
|  |  | (.087) |  |  | (.086) |
| Manager tenure |  |  | -.094** |  | -.055** |
|  |  |  | (.041) |  | (.045) |
| Manager educational level |  |  |  | -.642*** | -.572*** |
|  |  |  |  | (.189) | (.208) |
| Manager output experience | .07 | .067 | .078 | .057 | .063 |
|  | (.069) | (.07) | (.07) | (.07) | (.069) |
| Manager financial distress experience | -.005 | -.006 | -.006 | -.007 | -.006 |
|  | (.017) | (.017) | (.017) | (.017) | (.017) |
| Manager insolvency proceedings experience | .084 | .06 | .068 | .06 | .087 |
|  | (.07) | (.071) | (.07) | (.071) | (.07) |
| Manager founder | .083 | .011 | .085 | -.008 | .094 |
|  | (.08) | (.078) | (.084) | (.079) | (.084) |
| Firm age | -.001 | -.003 | -.002 | -.003* | -.001 |
|  | (.002) | (.002) | (.002) | (.002) | (.002) |
| Firm size (number of employees) | .05* | .042 | .039 | .044 | .051* |
|  | (.03) | (.031) | (.03) | (.03) | (.03) |
| Turnover t-1 | -.081*** | -.084*** | -.079*** | -.083*** | -.077*** |
|  | (.018) | (.018) | (.018) | (.018) | (.018) |
| Financial debts/ Assets t-1 | .001 | .003 | .004 | 0 | -.001 |
|  | (.022) | (.023) | (.022) | (.023) | (.022) |
| Cash and equivalent / Assets t-1 | .02 | .032 | .019 | .026 | .006 |
|  | (.046) | (.047) | (.047) | (.047) | (.046) |
| Firm Industry | -.019 | -.021 | -.023 | -.018 | -.016 |
|  | (.015) | (.015) | (.015) | (.015) | (.015) |
| _Cons | 4.849* | 2.481* | 2.583* | 2.748* | 4.995* |
|  | (.748) | (.292) | (.289) | (.337) | (.827) |
| R-squared | .158 | .127 | .14 | .132 | .17 |
| Maximum VIF | 2.00 | 2.00 | 2.00 | 2.01 | 2.02 |
| Observations | 342 | 342 | 342 | 342 | 342 |

Notes: The standard errors are shown in parentheses. The coefficients are estimated using a linear regression model. The maximum VIF represents the highest value of the Variance Inflation Factor among the independent variables. Asterisks indicate statistical significance: * $p<.1$, ** $p<.05$, *** $p<.01$.